\documentstyle[aps,prl,epsf]{revtex}

\newcommand{\be}{\begin{equation}}
\newcommand{\ee}{\end{equation}}
\newcommand{\bea}{\begin{eqnarray}}
\newcommand{\eea}{\end{eqnarray}}
\newcommand{\bdm}{\begin{displaymath}}
\newcommand{\edm}{\end{displaymath}}

\newcommand{\diff}{\mbox{$\rm{d}$}}
\newcommand{\Diff}{\mbox{$\rm{D}$}}
\newcommand{\cod}{\diff^{\dagger}}
\newcommand{\we}{\wedge}

\newcommand{\ubold}{\mbox{\boldmath $u$}}
\newcommand{\gtens}{\mbox{\boldmath $g$}}
\newcommand{\gtiltens}{\tilde{\gtens}}

\newcommand{\ghattens}{\hat{\gtens}}

\newcommand{\Rtens}{\mbox{\boldmath $R$}}

\newcommand{\nabhat}{\hat{\nabla}}

\newcommand{\hatast}{\hat{\ast}}
\newcommand{\barast}{\bar{\ast}}

\def\boxdal{\hbox{\hskip 0.5mm\hbox{\vrule width2.3mm height0.2mm
\vbox{\hrule width0.3mm height2.6mm}\hskip
-2.6mm \vbox{\hbox{\vrule width2.6mm height0.1mm}
\vskip -0.1mm\hrule width0.1mm height2.6mm}}\hskip 0.5mm}}

\begin{document}

\twocolumn[\hsize\textwidth\columnwidth\hsize\csname @twocolumnfalse\endcsname

\title{The generalization of the Regge-Wheeler equation for self-gravitating 
matter fields}

\author{O. Brodbeck$^{\star}$, 
        M. Heusler$^{\dagger}$, and O. Sarbach$^{\dagger}$}

\address{$^{\star}$Max-Planck-Institute for Physics, Werner Heisenberg
         Institute, D--80805 Munich, Germany\\
         $^{\dagger}$Institute for Theoretical Physics, University of Zurich,
         CH--8057 Zurich, Switzerland}

\maketitle

\begin{abstract}
It is shown that the dynamical evolution of perturbations 
on a static spacetime is governed by a standard pulsation 
equation for the extrinsic curvature tensor. The centerpiece 
of the pulsation equation is a wave operator whose spatial 
part is manifestly self-adjoint. In contrast to metric 
formulations, the curvature-based approach to gravitational 
perturbation theory generalizes in a natural way to 
self-gravitating matter fields. For a certain relevant subspace 
of perturbations the pulsation operator is symmetric with 
respect to a positive inner product and therefore allows
spectral theory to be applied. In particular, this is the 
case for odd-parity perturbations of spherically symmetric 
background configurations. As an example, the pulsation 
equations for self-gravitating, non-Abelian gauge fields 
are explicitly shown to be symmetric in the gravitational, 
the Yang Mills, and the off-diagonal sector.
(PACS numbers: 04.25.Nx, 04.40.-b, 04.70.Bw)\\
\end{abstract}
]

In general relativity, several interesting and even some of 
the outstanding problems lie, at least partially, within the scope 
of gravitational perturbation theory. A prominent example in 
vacuum gravity is the collision of black holes, for which the
close limit approximation to head-on collisions has proven to 
be in remarkable agreement with fully numerical simulations 
\cite{APPSS95}, \cite{PP94}. In the presence of matter, the 
stability of neutron stars with respect to non-spherical 
perturbations \cite{FM97}, the study of critical phenomena in 
the weakly non-spherical gravitational collapse \cite{MC}, 
as well as stability and local uniqueness questions for hairy 
black-holes and self-gravitating solitons \cite{BHSV}, 
\cite{SHB99} provide further topic examples wherein perturbation 
theory finds a natural application.

For vacuum gravity, perturbation theory is highly developed and
even for self-gravitating systems, important general properties
of the perturbation equations were established. In particular, 
it was recently shown that a manifestly hyperbolic, 
gauge-invariant formulation of gravitational perturbation theory
does exist for static background configurations \cite{CB}, 
\cite{AA98}. However, only for special systems with linear 
matter models, such as vacuum gravity and the Einstein-Maxwell 
system \cite{VM}, formulations of perturbation theory are known 
for which, at least on a static background, the perturbations 
are described by a system of {\it pulsation\/} equations, that 
is, by a wave operator whose spatial part is (formally) 
{\it self-adjoint\/}. In view of the powerful results for self-adjoint 
elliptic operators, such symmetric formulations of gravitational 
perturbation theory would be most valuable for analytical as 
well as numerical investigations of the problems above-mentioned, 
in particular for spherically symmetric background configurations.

In this letter we show how to cast the equations governing the
dynamical perturbations of static background configurations
into a system of pulsation equations. We do so by first
following ideas introduced by Choquet-Bruhat {\it et al.} 
for full, non-linear general relativity \cite{CB}. In a 
second step we show that the resulting linearized wave 
equation for the extrinsic curvature can be {\it symmetrized}.
The relevant inner product is positive definite on the 
subspace comprising all traceless perturbations. In contrast 
to the traditional metric approach to gravitational 
perturbation theory, the method introduced in this letter 
admits a natural generalization to non-vacuum perturbations 
of non-vacuum configurations. In order to illustrate why, in 
general, the metric approach fails to yield a self-adjoint
system of wave equations, we start with a brief account of the 
gauge-invariant formalism introduced by Gerlach and 
Sengupta \cite{GS}. For simplicity we restrict ourselves to 
odd parity perturbations.

\noindent
{\it The gauge-invariant metric approach --}
The odd parity perturbations of a spherically symmetric 
spacetime $(M,\gtens)$ with metric
$\gtens = \gtiltens + R^{2} \ghattens$
are parametrized in terms of a scalar field
$\kappa(x^{b})$ and a one-form
$h = h_{a}(x^{b}) \diff x^{a}$
on the two-dimensional pseudo-Riemannian
orbit space $\tilde{M} \equiv M / \mbox{SO(3)}$;
\be
\delta g_{ab} = 0 , \; \; \;
\delta g_{Ab} = h_{b} S_{A} , \; \; \;
\delta g_{AB} = 2 \kappa \, \nabhat_{(A} S_{B)},
\label{Eq-1}
\ee
where
$S_{A} \equiv (\hatast \diff Y^{\ell m})_{A}$ denote the 
transverse spherical vector harmonics and 
$2\nabhat_{(A} S_{B)} \equiv \nabhat_{A} S_{B} +\nabhat_{B} S_{A}$.
[Lower-case Latin indices $(a=0,1)$ refer to coordinates  
on $(\tilde{M}, \gtiltens)$, while capital Latin indices
$(A=2,3)$ refer to the standard coordinates on the metric 
sphere $(S^{2}, \ghattens)$]. Under infinitesimal coordinate 
transformations induced by a vector field $X^{\mu}$, the 
metric perturbations transform according to
$\kappa \rightarrow  \kappa + \varphi$ and
$h \rightarrow h + R^{2} \diff \left( R^{-2} \varphi \right)$,
where $\varphi$ is the scalar field parametrizing the odd parity 
transformations, $X^{a} = 0$, $X^{A} = \varphi \, g^{AB}S_B$.
In terms of the manifestly gauge-invariant one-form $H$,
\be
H \equiv h - R^{2}  \diff ( \frac{\kappa}{R^{2}} ),
\label{Eq-2}
\ee
the gauge-invariant components of the Einstein
tensor become $\delta G_{ab}^{inv} = 0$,
$\delta G_{AB}^{inv} = - \cod H \nabhat_{(A} S_{B)}$
for $\ell > 1$, and
\be
\delta G_{Ab}^{inv} \diff x^{b}  =  \frac{S_{A}}{2 R^{2}}
\left\{ \cod \left[ R^{4} \diff \left( \frac{H}{R^2}
\right) \right] + \lambda H \right\}.
\label{Eq-3}
\ee
Here, $\lambda \equiv (\ell-1)(\ell+2)$ and
$\cod = \tilde{\ast} \diff \tilde{\ast}$ denotes the
co-differential operator with respect to $\gtiltens$.

The vacuum equations, $\delta G_{\mu \nu}^{inv} =0$,
imply a wave equation for a single scalar field: 
The Regge-Wheeler (RW) equation \cite{RW57} for the radial
component of $H$ is derived by using the integrability
condition $\cod H = 0$ (if $\ell \neq 1$) to eliminate the
temporal component of $H$. Alternatively, one obtains a 
{\it covariant\/} derivation (which is not limited to stationary 
background configurations) by applying the operator 
$\tilde{\ast} \diff$ to Eq. (\ref{Eq-3}). This yields
\be
\left[ \tilde{\boxdal} + R \tilde{\boxdal} 
\left( \frac{1}{R}\right) +
\frac{\lambda}{R^2} \right] \Psi = 0,
\label{Eq-4}
\ee
where the scalar $\Psi \equiv R^3 \tilde{\ast} \diff (H/R^2)$ 
is gauge-invariant for all values of $\ell$, and
$\tilde{\boxdal}$ denotes the d'Alembertian with respect to
$\gtiltens$.

In the presence of matter fields the gauge-invariant 
Einstein tensor $\delta G^{inv}_{\mu \nu}$ is still adapted 
for analyzing stationary perturbations \cite{SHB99}. However, 
following the above lines, it turns out that the 
{\it evolution\/} equations for metric perturbations
cannot be cast into a {\it self-adjoint\/} form.
In fact, as the derivation of the Teukolsky equation suggests, 
it cannot be expected that the gravitational perturbations are 
parametrized in terms of a single scalar amplitude, unless for
vacuum (or electro-vacuum) perturbations on a algebraically 
special background. It is, therefore, natural to look for a
wave equation for the one-form $H$ itself. However, the 
operator governing the evolution of $H$ is not hyperbolic, 
since the expression (\ref{Eq-3}) comprises only ``one half'' 
of the d'Alembertian. Hence, we are aiming at a
description of gravitational perturbations for which the
wave operator already appears ``off-shell''. Laying stress
on the {\it staticity} rather than the spherical symmetry of
the background, the natural gauge-invariant perturbations
turn out to be curvature quantities.

\noindent
{\it The gauge-invariant curvature-based approach --}
Under infinitesimal coordinate transformations generated by
a vector field $(X^{\mu}) = (f,X^{i})$, the metric perturbations
of a {\it static} background,
$\gtens = -\alpha^{2} \diff t^{2} + \bar{\gtens}$,
transform according to
$\delta \alpha  \rightarrow  \delta \alpha + \alpha_{,i} X^i +
\alpha \dot{f}$,
$\delta g_{ti}  \rightarrow  \delta g_{ti} + \dot{X}_{i} -
\alpha^2 f_{,i}$ and
$\delta g_{ij}  \rightarrow  \delta g_{ij} + X_{i;j} + X_{j;i}$,
where a dot denotes differentiation with respect to $t$, and
covariant derivatives refer to the Riemannian background 
metric $\bar{\gtens}$. Hence, the tensor
$\delta \dot{g}_{ij} - \delta g_{ti;j} - \delta g_{tj;i}$
is gauge-invariant under transformations with $f=0$.
Considering odd parity perturbations on a spherically 
symmetric background from this point of view, the natural 
gauge-invariant amplitudes are given by the $(1A)$ and 
the $(AB)$ component of the above tensor. By virtue of 
Eqs. (\ref{Eq-1}) these are proportional to
\be
\dot{h}_{1} - R^{2} \partial_1 \left( \frac{h_{t}}{R^{2}} \right) ,
\; \; \;
\dot{\kappa} - h_{t} .
\label{Eq-6}
\ee
While the RW one-form $H$ is suited to describe stationary 
perturbations, the gauge-invariant variables (\ref{Eq-6}) 
are adapted to their time-evolution. Note that the situation 
is reminiscent of Maxwell theory, where the gauge-invariant 
quantity satisfying a wave equation is the electric field, 
$E_i \equiv - \dot{A}_i + \Phi,_i$,
which is constructed in a similar way as the tensor 
$\delta \dot{g}_{ij} - \delta g_{ti;j} - \delta g_{tj;i}$.
Since the latter essentially is the perturbation of the extrinsic 
curvature $K_{ij}$, our goal is to derive a symmetric wave 
equation for $\delta K_{ij}$. Within the curvature-based 
approach the RW equation has, as we shall show below,
a {\it two\/}-component analogue which admits generalizations 
to self-gravitating field theories in a straightforward way.

\noindent
{\it The symmetric wave equation for the extrinsic curvature --}
In the Arnowitt-Deser-Misner (ADM) formalism the metric reads
\be
\gtens = -\alpha^{2} \diff t^{2} +
\bar{g}_{ij} (\diff x^{i} + \beta^{i} \diff t)
(\diff x^{j} + \beta^{j} \diff t) ,
\label{Eq-7}
\ee
where the shift $\beta^i$ and the extrinsic curvature
$K_{ij} \equiv (2\alpha)^{-1}
\left([\partial_{t} - {\cal L}_{\beta}] \bar{\gtens}\right)_{ij}$
vanish on a static background. An efficient way to linearize 
the ADM equations is to use ``vector-invariant'' quantities, 
that is, perturbations which are invariant under the subset 
of transformations generated by vector fields 
$(X^{\mu}) = (f,X^i)$ with $f=0$. These quantities are
$\delta R_{0i}$, $\delta K_{ij}$, and
\be
\delta \dot{\alpha} - {\cal L}_{\delta \beta} \alpha,
\; \; \;
\delta\dot{\bar{g}}_{ij} - 
\left({\cal L}_{\delta \beta} \bar{\gtens} \right)_{ij},
\; \; \;
\delta \dot{R}_{ij} - \left({\cal L}_{\delta \beta} \Rtens \right)_{ij},
\label{Eq-8}
\ee
where the index ``0'' refers to the normal unit vector field 
$\partial_{0} = \alpha^{-1} (\partial_{t} - \beta^{i}\partial_{i})$.
As the variation of the shift enters the ADM equations via 
the combinations (\ref{Eq-8}) only, one may set $\delta \beta = 0$ 
for computations, provided that in the resulting expressions
all perturbations are identified with their 
vector-invariant counterparts. The only gauge freedom is then 
parametrized by the function $f$. In particular, one has
\be
\left[ \alpha^2 \delta K - \delta \dot{\alpha} \right]
\rightarrow
\left[ \alpha^2 \delta K - \delta \dot{\alpha} \right] -
\alpha \ddot{f} + \alpha^2 \left(\alpha f^{,i} \right)_{;i},
\label{Eq-9a}
\ee
where now $\delta \dot{\alpha}$ denotes the
vector-invariant lapse defined in (\ref{Eq-8}).
By virtue of this transformation law the scalar gauge
freedom can be used to impose the harmonic gauge,
$\alpha^2 \delta K = \delta \dot{\alpha}$.
Adopting this gauge and using vector-invariant quantities 
removes all gauge degrees of freedom, up to residual
gauge transformations generated by $(X^{\mu})=(f,0)$ with
$\ddot{f}=\alpha (\alpha f^{,i})_{;i}$. The
linearized ADM evolution equations then become
\be
\delta K_{ij} = \frac{1}{2 \alpha} \delta \dot{\bar{g}}_{ij} , \; \; \;
\delta R_{ij} = \delta \bar{R}_{ij} +
\frac{\delta \dot{K}_{ij}}{\alpha} -  \delta
\frac{\bar{\nabla}_{i}\bar{\nabla}_{j} \alpha}{\alpha} ,
\label{Eq-9}
\ee
while the linearized constraints are given by
\be
\delta G_{00} = \frac{1}{2} \delta \bar{R}, \; \; \;
\delta R_{0i} = \bar{\nabla}^{j} \delta K_{ij} -
\bar{\nabla}_{i} \delta K .
\label{Eq-10}
\ee

In order to obtain a wave equation for $\delta K_{ij}$ we first 
follow the method introduced by Choquet-Bruhat {\it et al.} 
\cite{CB}, and then show how to cast their result into a 
symmetric form. More precisely, one may proceed as follows:
First, one differentiates the second evolution equation
in Eqs. (\ref{Eq-9}) and uses the first one to obtain an 
expression 
for $\delta \dot{R}_{ij}$ in terms of $\delta \ddot{K}_{ij}$
and covariant derivatives of $\delta K_{ij}$ up to
second order. Second, one eliminates the variation of the 
lapse by adopting the harmonic gauge,
$\delta \dot{\alpha} = \alpha^2 \delta K$.
Next, one gets rid of those second covariant
derivatives of $\delta K_{ij}$ which are not of the
desired form by subtracting
$\alpha[(\delta R_{0i})_{;j} + (\delta R_{0j})_{;i}]$ from
$\delta \dot{R}_{ij}$ and applying
the commutation law for covariant derivatives.
The resulting expression,
which still contains first covariant derivatives
of $\delta K_{ij}$, is finally symmetrized by subtracting
$3 [\alpha_{;j} \delta R_{0i} + \alpha_{;i} \delta R_{0j}]$.
The result is that the tensor
\be
\Lambda_{ij} \equiv \alpha^{2} \partial_t
\delta R_{ij} - 
\left(\alpha^{3} \delta R_{0i} \right)_{;j} -
\left(\alpha^{3} \delta R_{0j} \right)_{;i}
\label{Eq-11}
\ee
yields a formally self-adjoint, hyperbolic operator
acting on $\delta K_{ij}$. [The difference between the
$\Lambda_{ij}$ defined in Eq. (\ref{Eq-11}) and the 
$\Omega_{ij}$ introduced in
Ref. \cite{CB} lies in the symmetrizing terms.] One finds
\bea
\Lambda_{ij} & = &
\boxdal \delta K_{ij} +
2\alpha^3\bar{R}^{l}_{(i} \delta K_{j)l} - 2\alpha^3\bar{R}_{likj}
\delta K^{kl}
\nonumber\\ & & +
4 \, \alpha\alpha^k \bar{\nabla}_{(i}\left[ \alpha\delta K_{j)k} \right] -
4 \, \alpha\alpha_{(i} \bar{\nabla}^k\left[ \alpha\delta K_{j)k} \right]
\nonumber\\ & & +
2\bar{\nabla}_{(i}\left[ \alpha\alpha^k \right]
\alpha\delta K_{j)k} -
2\bar{\nabla}_{i}\left[ \alpha\alpha_j \right]  \alpha 
\bar{g}^{kl} \delta K_{kl},
\label{Eq-12}
\eea
where $\boxdal \delta K_{ij}  \equiv
[\alpha \partial_t^2 - \alpha \bar{\nabla}^k \alpha
\bar{\nabla}_k \alpha] \delta K_{ij}$.
The complete operator in Eq. (\ref{Eq-12}) acting on the 
linearized extrinsic curvature is symmetric with respect to 
the inner product induced by the De Witt metric
\be
G^{ijkl} \equiv
\bar{g}^{ik}\bar{g}^{jl} - \bar{g}^{ij}\bar{g}^{kl}.
\label{Eq-13}
\ee
Hence, the inner product is positive definite only for perturbations
with vanishing trace, $\delta K = 0$. In particular, this is 
the case for odd parity perturbations of spherically symmetric 
background configurations.

For a spherically symmetric background with metric 
$\bar{\gtens} = \alpha^2 \diff \rho^2 + R^2 \diff \Omega^2$,
the multipole expansions of the linearized vacuum equations,
$\Lambda_{\rho A} = 0$, $\Lambda_{AB}=0$, yield the pulsation 
equation $[\tilde{\boxdal} + P] \ubold = 0$ and the (momentum) 
constraint $(R^2 \alpha^{-1} u_1)' = \sqrt{\lambda} R u_2$, where
\bdm
P \equiv \frac{1}{R^2 \alpha^2} 
\left(\begin{array}{cc}
R^4/\alpha (\alpha/R^{2})'' + \lambda \alpha^2 &
2 \sqrt{\lambda} R^2 (\alpha/R)' \\
2 \sqrt{\lambda} R^2 (\alpha/R)' &
R R'' + \lambda \alpha^2
\end{array}\right).
\edm
The gauge-invariant amplitudes $\ubold = (u_1, u_2)$
are defined by $\delta K_{\rho B} \equiv u_1 S_B$ and
$\delta K_{AB} \equiv R \alpha^{-1} \lambda^{-1/2} u_2 
\hat{\nabla}_{(A}S_{B)}$, 
the d'Alembertian refers to the background metric
$\tilde{\gtens} = \alpha^2(-\diff t^2 + \diff \rho^2)$,
and a prime denotes differentiation with respect to $\rho$.

A convenient way to solve the initial value problem is to
impose initial conditions
$\ubold(0,\rho)$ and $\dot{\ubold}(0,\rho)$
subject to the constraint equation and its time derivative.
Having solved the wave equation for $\ubold(t,\rho)$, the 
metric perturbations are then obtained from the $\delta R_{ij}$ 
equation by using the relations 
$-2 u_2= \sqrt{\lambda} R^{-1} H_{t}$ and
$-2 u_1 = \alpha R^2 \tilde{\ast} \diff (R^{-2} H)$
between the old and the new gauge-invariant amplitudes.

\noindent
{\it Pulsation equations for self-gravitating 
matter fields --}
With respect to the ADM basis of one-forms the
gauge potential $A$ is parametrized by a scalar 
field $\Phi$ and a one-form $\bar{A}$ (both Lie algebra valued),
\be
A = \Phi \, \alpha \diff t + \bar{A}_{i}
\left(\diff x^{i} + \beta^{i} \diff t \right) .
\label{YM-1}
\ee
In order to linearize the YM equations we
proceed in a similar way as in the gravitational
case. While the fundamental gravitational object is
the extrinsic curvature,
$K_{ij} \equiv \frac{1}{2} {\cal L}_{\partial_{0}} g_{ij}$,
the corresponding quantity in the YM case is
the electric one-form $E = -i_{\partial_{0}} F$.
Here, $F = \diff A + A \we A$ is the YM field strength and
$i_{\partial_{0}}$ denotes the inner derivative
with respect to the normal unit vector field $\partial_{0}$.
For a static, purely magnetic background one has
$A = \bar{A}$, while $\Phi$ and $E$ vanish.
The linearized electric field thus becomes
\be
\alpha \delta E = \bar{\Diff}(\alpha \delta \Phi) -
\delta \dot{\bar{A}} +
{\cal L}_{\delta \beta} \bar{A},
\label{YM-2}
\ee
where $\bar{\Diff} = \bar{\diff} + [\bar{A},\;]$
is the three-dimensional gauge covariant exterior derivative.
Since $\delta A \rightarrow \delta A + {\cal L}_{X} A$
under coordinate transformations, $\delta E$, $\delta \Phi$ and
$\delta \dot{\bar{A}} - {\cal L}_{\delta \beta} \bar{A}$
are vector-invariant quantities. Hence, the variation of the shift 
can be taken into account by using vector-invariant metric 
and matter perturbations. It then remains to linearize  the YM 
equations with respect to the time-dependent metric
$\gtens = -\alpha^{2} \diff t^{2} + \bar{g}_{ij}
\diff x^{i} \diff x^{j}$ and the YM potential
$A = \Phi \alpha \diff t + \bar{A}$. As $\Phi$ and the temporal
derivatives of all background quantities vanish, the evolution 
equations become
\be
\delta \dot{\bar{A}} = \bar{\Diff}
\left(\alpha \delta \Phi \right) - \alpha \delta E , \; \; \;
\delta \dot{E} = \delta \bar{\Diff}^{\dagger}
\left( \alpha \bar{F} \right) ,
\label{YM-3}
\ee
while the Gauss constraint is given by
\be
\bar{\Diff}^{\dagger} \delta E = 0 ,
\label{YM-4}
\ee
where, for Lie algebra valued $p$-forms,
$\bar{\Diff}^{\dagger} \equiv (-1)^{p}
\barast \bar{\Diff} \barast$.
These equations hold for
arbitrary variations of the shift, provided that
all perturbations are identified with their vector-invariant
counterparts.
Moreover, $\delta E$ is invariant under arbitrary
gauge transformations generated by Lie algebra valued
scalar fields $\chi$, since
$\delta \Phi \rightarrow \delta \Phi + \dot{\chi}$
and $\delta \bar{A} \rightarrow \delta \bar{A} + \bar{\Diff} \chi$.

The linearized gravitational equations (\ref{Eq-9}), (\ref{Eq-10}) 
and the YM equations (\ref{YM-3}), (\ref{YM-4}) have the same 
structure, where  $\delta K_{ij}$ and $\delta E$, as well as
$\delta \dot{\bar{g}}_{ij}$ and $\delta \dot{\bar{A}}$
correspond to each other. In both cases one has to differentiate the
second evolution equation and to use the first one to obtain 
hyperbolic equations for $\delta K_{ij}$ and $\delta E$. 
Also in both situations one needs to complement the resulting
expressions with spatial derivatives of the constraints to obtain 
the desired wave operator. Performing this procedure for the YM equations
(\ref{YM-3}), (\ref{YM-4}), and using the background
equation $\bar{\Diff}^{\dagger} ( \alpha \bar{F} ) = 0$
yields the linearized YM equation
\be
\Lambda = \alpha \partial_t 
\left[ \delta_{{\mbox{metric}}} \bar{\Diff}^{\dagger} 
\left(\alpha \bar{F}\right) \right].
\label{YM-RHS}
\ee
The RHS describes the coupling to the metric perturbations
$\delta \alpha$ and $\delta \bar{g}_{ij}$, while the
one-form $\Lambda$ comprises the matter perturbations;
\be
\Lambda \equiv \boxdal  \delta E +
\alpha^{3} \barast [\delta E , \barast \bar{F} ],
\label{YM-wave}
\ee
where
$\boxdal \delta E \equiv
[\alpha \partial^2_t + 
\bar{\Diff} \alpha^{3} \bar{\Diff}^{\dagger} +
\alpha \bar{\Diff}^{\dagger} \alpha \bar{\Diff} \alpha ]
\delta E$. Both, the wave operator and the potential in 
Eq. (\ref{YM-wave}) are symmetric with respect to the invariant
inner product 
$-4G\int g^{ij} \, \mbox{trace} 
\left\{ \delta E^{(1)}_i \delta E^{(2)}_j \right\}$.

It is remarkable that the linearized gravitational equation
\be
\Lambda_{ij} = 8\pi G \left[
\alpha^{2} \partial_t \delta T_{ij} -
\left(\alpha^{3} \delta T_{0i} \right)_{;j} - 
\left(\alpha^{3} \delta T_{0j} \right)_{;i} \right],
\label{E-RHS}
\ee 
and the linearized YM equation (\ref{YM-RHS}),
with $\Lambda_{ij}$ and $\Lambda$ given in Eqs. (\ref{Eq-12})
and (\ref{YM-wave}), respectively, perfectly fit together:
All matter perturbations arising on the
RHS of the gravitational equation (\ref{E-RHS})
can be expressed in terms of $\delta E$ by using
the first YM evolution equation in (\ref{YM-3}) and its
first spatial derivative. In a similar way, the RHS of the YM
equation (\ref{YM-RHS}) involves only metric perturbations
which can be expressed in terms of $\delta K_{ij}$ by taking
advantage of the first ADM evolution equation in
(\ref{Eq-9}). Hence, in the wave equation for 
$\delta K_{ij}$ the matter perturbations arise via $\delta E$
only, and vice-versa. Moreover, using the momentum constraint,
the off-diagonal contributions arising from the coupling between 
the gravitational and the YM sector enter Eqs. (\ref{YM-RHS}) 
and (\ref{E-RHS}) in a perfectly symmetric way. We thus end up with
a single, formally self-adjoint wave equation describing both
gravitational and matter field perturbations.

The derivation of the RHS of Eqs. (\ref{YM-RHS}) and (\ref{E-RHS})
will be given elsewhere \cite{longversion}. Here we only present 
the result: The gravitational pulsation equation (\ref{E-RHS})
becomes
\bea
\Lambda_{ij} & = & 8 G \, \mbox{trace} \Big\{
\alpha^3 \left[ \bar{F}^k_{\;\, i} \bar{F}^{l}_{\;j} -
\frac{1}{2}\bar{g}_{ij}\bar{F}^k_{\;m}
\bar{F}^{l m} \right] \delta K_{k l} 
\nonumber\\ & & + \,
\frac{\alpha^3}{4} \bar{F}_{k l}\bar{F}^{k l} \delta K_{ij} -
\frac{1}{2}\bar{g}_{ij} \alpha^2\bar{F}^{k l} \bar{D}_k \left[
\alpha\delta E_{l} \right]
\nonumber\\ & & + \,
\bar{D}_k\left[ \alpha\delta E_{(i} \right]
\alpha^2\bar{F}^k_{\; \, j)} + \alpha\delta E_k 
\bar{D}_{(i}\left[ \alpha^2\bar{F}^k_{\;\, j)} \right]
\Big\}
\label{Eq-A}
\eea
with $\Lambda_{ij}$ according to Eq. (\ref{Eq-12}).
The linearized YM equation (\ref{YM-RHS}) is given by
\bea
\Lambda_i & = & -8 G \, \alpha^3 \bar{F}^k_{\;\, i} \, \mbox{trace}
\left\{ \delta E_j\bar{F}^j_{\; k} \right\} +
2\alpha \bar{D}^j \left[ \alpha^2\bar{F}^k_{\; \, i} \right]
\delta K_{jk}
\nonumber\\ & & -
2\alpha^2 \bar{F}^k_{\; \, j} \bar{\nabla}^j \left[
\alpha \delta K_{ik} \right]
+ 2\alpha^2 \bar{F}_{ij}\alpha^j \delta K
\label{Eq-B}
\eea
with $\Lambda_i$ according to Eq. (\ref{YM-wave}).
It is not hard to verify that the above equations form
a symmetric system with respect to the inner products
given above. The perturbations $\delta K_{ij}$ and $\delta E_i$
have to satisfy the momentum constraint and the YM
Gauss constraint (\ref{YM-4}), respectively, while
$\delta\bar{g}_{ij}$ and $\delta\bar{A}_i$ are subject to the
Hamilton constraint and Eqs. (\ref{Eq-9}) and (\ref{YM-3}).

For odd parity perturbations on a spherically symmetric background, 
the multipole decomposition of 
Eqs. (\ref{Eq-12}), (\ref{YM-wave}), (\ref{Eq-A}) and (\ref{Eq-B})
yields a similar pulsation equation as
in the vacuum case, where $\ubold$ now comprises two
gravitational and five YM amplitudes [for SU(2)]. 
The explicit form of this pulsation equation
as well as a gauge-invariant formulation of all
constraint equations will be given in Ref. \cite{longversion}.

\noindent
{\it Conclusion --}
The dynamical perturbations of a static spacetime are
governed by a system of standard pulsation equations for the
linearized extrinsic curvature. In the presence of matter fields,
the curvature-based gravitational perturbations are naturally 
embedded in a larger system of standard pulsation equations, 
comprising both gravitational and matter perturbations. 
In particular, the dynamical perturbations of static EYM 
configurations are described by a wave operator whose spatial 
part is manifestly self-adjoint. As the relevant inner product 
in the gravitational sector is positive definite only for traceless 
perturbations, further investigations are needed to discuss, 
for instance, the dynamics of even parity perturbations in a
satisfactory way. For perturbations with vanishing trace, in 
particular for odd parity perturbations of spherically symmetric
configurations, the form of the pulsation equation is optimal and 
enables one to apply the tools from spectral analysis.

\end{document}